\documentclass[conference]{IEEEtran}

\renewcommand\IEEEkeywordsname{Keywords}
\pdfoutput=1 

\IEEEoverridecommandlockouts
\usepackage{cite}
\usepackage{amsmath,amssymb,amsfonts}
\usepackage{algorithmic}
\usepackage{graphicx}
\usepackage{textcomp}
\usepackage{xcolor}
\def\BibTeX{{\rm B\kern-.05em{\sc i\kern-.025em b}\kern-.08em
    T\kern-.1667em\lower.7ex\hbox{E}\kern-.125emX}}
\begin{document}

\title{Effective Route Scheme of Multicast Probing to Locate High-loss Links in OpenFlow Networks\\
}

\author{\IEEEauthorblockN{Nguyen Minh Tri\IEEEauthorrefmark{1},
		Masahiro Shibata\IEEEauthorrefmark{1},
		Masato Tsuru\IEEEauthorrefmark{1}}
	\IEEEauthorblockA{\IEEEauthorrefmark{1}{School of Computer Science and Systems Engineering, Kyushu Institute of Technology, Japan}\\
		}
}

\maketitle

\begin{abstract}
With the prevalence of cloud computing and contents delivery networking, OpenFlow-based centrally-managed networks with flexible and dynamic traffic engineering are demanded. To maintain a high quality of service of the network, detecting and locating high-loss links is essential.
Therefore, in this paper, a measurement framework is proposed to promptly locate all high-loss links with a minimized load on both data-plane and control-plane incurred by the measurement, which assumes only standard OpenFlow functions.
It combines an active measurement by probing multicast packets along a designed route and a passive measurement by collecting flow-stats of the probing flow at selected switch ports in an appropriate sequential order to access switches.
In particular, by designing the measurement route based on the backbone-and-branch tree (BBT) route scheme, the measurement accuracy and the measurement overhead (i.e., the number of accesses to switches until locating all high-loss links) can be balanced.
The numerical simulation demonstrates the effectiveness of our proposal.
\end{abstract}


\section{Introduction}
The Software Defined Network (SDN) technology in general and OpenFlow in particular have attracted attention over the years \cite{SDN14}, and are spreading as a replacement solution for traditional network not only in data centers but also in enterprise networks and wide area networks, so called SD-WAN.
By decoupling the control plane and data plane, SDN becomes more flexible and simpler to design, configure, operate, and monitor. The deployment of services and features in the network is executed by programming on controllers. 
In data plane, switches forward packets based on rules from controllers. 
In particular, the ongoing prevalence of cloud computing and contents delivery networking requires a flexible traffic engineering on a network connecting globally-distributed datacenters, which is often centrally managed by OpenFlow \cite{Google13,MS13}.

Monitoring is an essential task in network management and operations. Operators need to know the network status/performance information in a real-time manner to make decisions about trouble-shooting, dynamic routing, load balancing, Service Level Agreement management, and so on. 
In general, there are two kinds of measurement approaches: passive and active. The passive approach is used to monitor link traffic state by collecting the statistical information (e.g., flow-stats) from switches (through the OpenFlow monitoring messages or SNMP) or by monitoring the OpenFlow-standard operating messages themselves.
There is a trade-off between the measurement accuracy and the measurement overhead in the control network.
Polling at a high frequency can increase the timeliness and accuracy, but it also causes more load incurred on switches and the control network.
There are some studies to reduce such a load. In OpenTM \cite{OpenTM}, each flow is measured by a periodical query to one switch. However, the switch decision can affect accuracy. In FlowSense \cite{FlowSense}, by only using FlowRemoved and PacketIn messages of OpenFlow standard, status on network utilization can be calculated with no additional cost, but it cannot trace quickly changed links. In PayLess \cite{PayLess}, a dynamic algorithm to balance the request frequency and the accuracy was introduced.
Similarly, in OpenNetMon \cite{OpenNetMon}, an adaptive polling rate to access edge switches was designed to reduce network and switch CPU overhead while optimizing the accuracy in throughput, delay and packet loss measurements.

On the other hand, the active approach sends and receives probe packets to measure the packet loss, delay, the round-trip-time (RTT), and so on. With the development of the edge-cloud computing for emerging IoT technologies, it is required reliable networks among a large number of heterogeneous sites over geographically-wider locations. In such networks, a ``link'' between two nodes is not always physical but sometimes virtual (e.g., tunneling) one that travers inaccessible intermediate switches prohibited from being monitored by passive measurements. Therefore an active measurement by probing packets is essential to monitor entire network information.
Furthermore, to realize a highly flexible and dynamic traffic engineering in OpenFlow networks, status/performance of all links should always be monitored and performance- deteriorated links should be located in a real-time manner.
However, probing at a high sending rate for a long duration can cause more load incurred on switches and the data network. Therefore, there are some studies to reduce such a load but still retain the reliability and precision. 
Adding an active measurement function (i.e., agent that sends and receives probe packets) on some or all switches that are globally controlled by a manager is a straight-forward approach and was implemented in some dedicated switches such as Cisco's Service Assurance Agent (SAA)/Internetwork Performance Monitor (IPM).
However this approach requires a special function beyond OpenFlow standard on each switches.
Authors in \cite{GRAMMI} proposed an infrastructure to monitor RTT; it focuses on reducing the flow entries and the number of probe packets. In \cite{Shibuya16}, a measurement scheme that can cover all links in both directions with minimizing flow entries on switches is presented. 
Focusing on datacenter networks, in \cite{Netbouncer}, a controller designs the probing routes and the multiple probing servers send probe packets along the designed routes, which are bounced by some switches back to the servers; then a processor collects the resulting data by accessing the servers. 
Such arbitrary (effective) probing routes are realized by IP-in-IP technique.
In \cite{deTector}, a real-time failure location is achieved by devising the design of effective probe matrix and using source routing of probe packets.
However, since they all use unicast probing in an end-to-end (among servers or beacons) manner, 
some links may suffer from many overlapped probing paths traversing them.

The Boolean (or performance) network-tomographic approaches have been studied, which only monitor performance-level correlations among measurement paths to infer the location of bad internal links. Seminal works such as \cite{Duffield06} have attracted much attention and been followed by a number of studies because of its practicality (e.g., \cite{Tachibana06}).
However, network-tomographic approaches always work with a considerable inference errors.
The impact of the capability of routing of probe packets has also been studied in localizing failed nodes based on Boolean network tomography \cite{LoNodeFail}.

In this paper, based on and motivated by those existing work, we present a network-assisted measurement framework for OpenFlow networks to monitor all links in both directions distinguishably to promptly and efficiently locate high-loss (i.e., performance-deteriorated) links, which aims to minimize the load on both data-plane and control-plane incurred by the measurement.
In contrast to existing works, our framework combines an active measurement by probing multicast packets from a measurement host to appropriate switch ports and a passive measurement by collecting flow-stats of the probing flow at selected switch ports in an appropriate sequential order to access switches that is determined dynamically.
The former reduces unnecessary load on the data plane incurred by probe packets and avoid the concentration at a link near the measurement host, while the latter reduces unnecessary load on the control plane incurred by switch accesses until all high-loss links are located.
Note that a simple Boolean network-tomographic inference of highly lossy range (a sequence of links) is used to dynamically determine a sequential access order to collect flow-stats.
In contrast to the typical network-tomographic approaches,
a tomographic approach in the proposed framework is not used for finally identifying of the lossy links;
it is used as a hint for narrowing the search space and for optimizing the search order.

This paper is an enhanced version of our preliminary conference paper \cite{TriISCC19} but includes a completely new route scheme that significantly outperforms the route scheme based on the shortest path tree in \cite{TriISCC19}, in terms of less number of accesses to switches to reduce the unnecessary load on the control-plane. 
In another extension by our group \cite{GotoICOIN20}, a link weight is introduced in the shortest path tree-based route scheme by using the results of past measurements, i.e. information on which links are likely to be lossy, in order to place loss-prone links near the ends (leaves) of a route tree. 
However, since it was also based on the shortest path tree, it should be improved by our new scheme in this paper.

The basic system model is presented in the next section. The probe packet route algorithm is present in Section 3. How to dynamically determine a sequential access order to collect flow-stats to efficiently locate/identify the high-loss links is shown in Section 4.
Section 5 discusses the design of the route scheme.
Section 6 provides the experimental results through simulation. 
The concluding remarks are given in the last section.

\section{System Overview}

\begin{figure}[t]
	\centering{\includegraphics[width=0.9\linewidth]{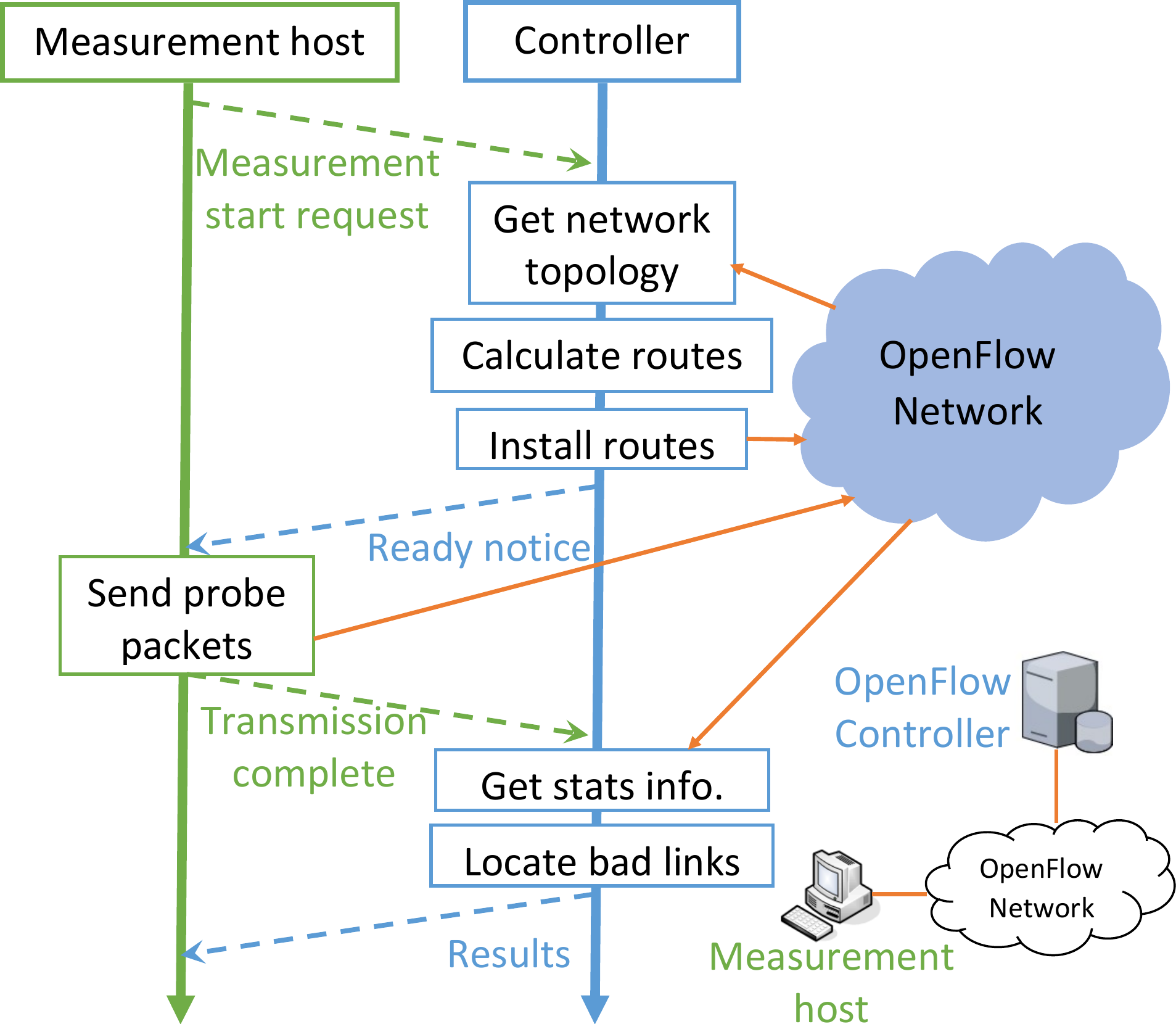}}
	\caption{Measurement process to locate bad links \cite{TriISCC19}}
	\label{process}
\end{figure}

\begin{figure}[t]
	\centering{\includegraphics[width=0.99\linewidth]{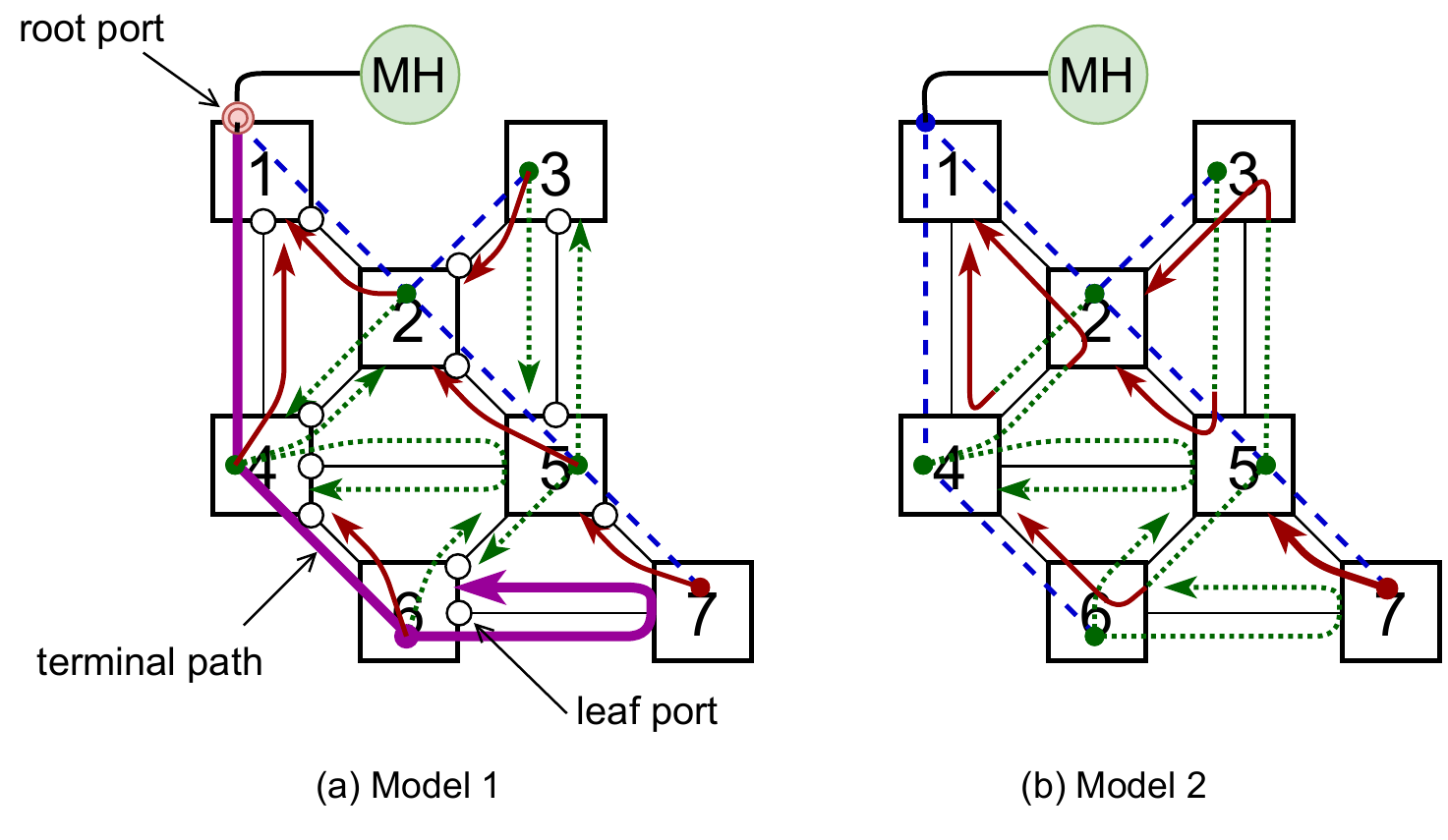}}
	\caption{Route scheme in \cite{TriISCC19}}
	\label{p_route}
\end{figure}

The proposed method is based on the framework that we previously proposed to monitor and locate high-loss links using multicast probing on OpenFlow networks \cite{TriISCC19}. It assumes the standard functions of OpenFlow-based networks comprising OpenFlow controller (OFC) and OpenFlow switches (OFS), i.e., the target networks include per-flow flexible routing/multicasting and per-flow monitoring of network statistics in a centralized manner; and is implemented on the OFC. The process begins when the measurement host (MH) sends a measurement request to the OFC, as in Fig. \ref{process}. Then, the OFC obtains network topology, calculates probe packet routes, and installs them to OFSs. 
Fig. \ref{p_route} shows an example of route scheme that are explained later in Section 3.1. The switch port connected to the MH is the root port; the leaf port is a switch port which discards probe packets. A route of the probe packets (i.e., the measurement flow) from the root port to a leaf port is referred to as a terminal path. The number of links from the root port to the leaf port is the path length.

Following that, a series of probe packets is launched by a single MH. Here, each probe packet (or a copy) passes through each link once and only once (in each direction of a full-duplex link separately) and is discarded at a leaf port on the last OFS along the terminal path.
The number of probe packets arriving at an individual input port on each OFS is recorded and collected by the OFC if required. Then, the packet loss rate on a link (or a sequence of links) between two switch ports is calculated by taking the difference between the numbers of arriving probe packets at those two ports.

Two important features that are strongly correlated are (i) flexible design of multicast measurement on a route tree with an MH location (i.e., the root of the tree) to cover all links in the active probing and (ii) dynamic optimization of the sequential access order to switches for collecting the flow-stats of the measurement flow at some switch ports that are passively monitored to locate high-loss links. 
In \cite{TriISCC19}, a shorted-path tree based multicast route scheme is used, where each probe packet traverses each link only once to minimize unnecessary load on the data-plane in OFS incurred by probe packets. This can avoid concentration of probe packets at links near the MH, especially in large networks. 
Different possible multicast measurement routes (including a single unicursal unicast measurement route over all links as an extreme case) can have the same benefit relative to data-plane load; however, measurement robustness and accuracy are strongly affected by the measurement route. For example, when a large number of probe packets is lost on a given link, all succeeding downstream links on that terminal path may not be monitored accurately due to a reduced number of probe packets passing through those links. Thus, a very long single unicursal route should not be used.

To reduce the time required and the unnecessary load on the control plane in the OFC and OFSs to locate all high-loss links, a sequential access order for the necessary flow-stats on the required OFSs is determined dynamically.
However, minimizing the length of each terminal path creates many short terminal paths, especially in large-scale networks, and results in more accesses to OFSs from the OFC.
This is because at least the OFC needs to collect the flow-stats at the root port and at every leaf ports to get the loss rates of all terminal paths.

Therefore, in this paper, we propose a new route scheme that can balance between the measurement accuracy and the measurement overhead (i.e., the number of accesses). Our proposal can keep the terminal paths short enough to avoid errors in loss rate estimation, while keeping the number of terminal paths small enough to reduce the necessary number of accesses to switches significantly.

\section{Route Scheme Design}

\begin{figure}[t]
	\centering{\includegraphics[width=0.7\linewidth]{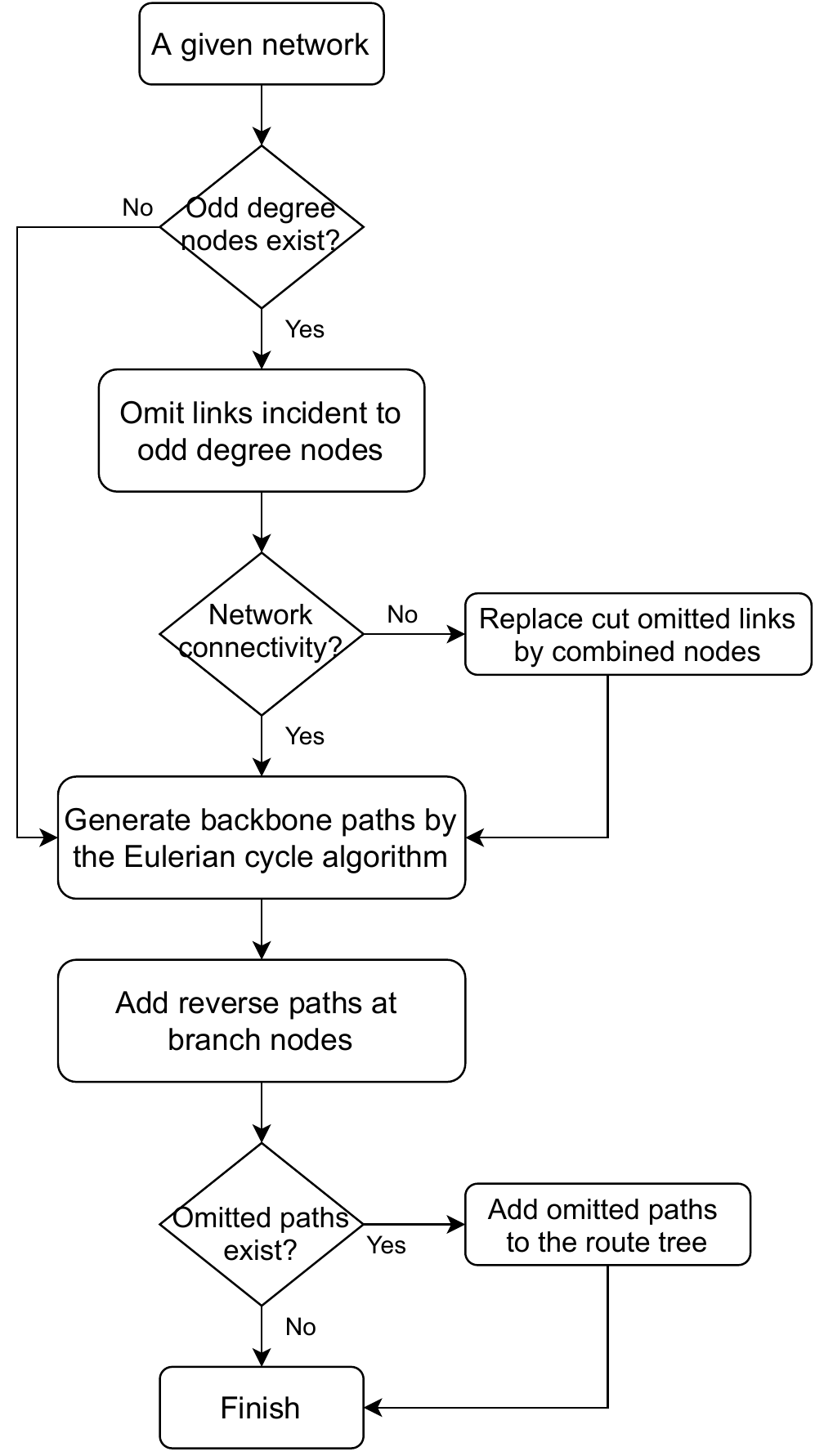}}
	\caption{Proposed route tree design flowchart}
	\label{flowchart}
\end{figure}

\subsection{Two baseline route schemes}
The measurement route traversed by probe packets strongly impacts on the search performance, i.e., how it can locate all high-loss links with a small load. To minimize the load on the data-plane, probe packets have to travel every link only once.
Please note that each link is assumed to be in the full-duplex mode and the probe packets should traverse in each direction on every link.
To fulfill this requirement, a simplest measurement route is a unicast trail.
Since link is full-duplex, each undirected link between OFSs is considered as two oppositely directed edges and OFSs are considered as vertices. In such a directed graph, from any vertex, the Eulerian cycle (circuit) algorithm can find a trail that passes every edge exactly once because each vertex has an even number of edges (i.e., it has an even degree).
We call this trail is the unicursal path. Its length is the double of the number of links. In this case, the route scheme has only one terminal path with the maximum length.

However, as discussed above, a long terminal path is not a good choice. In \cite{TriISCC19}, we proposed a route scheme with minimum length paths, called Model 1 as Fig. \ref{p_route}a, and a variant from it, Model 2, with longer paths, asFig. \ref{p_route}b. Both models use Dijkstra's shortest path algorithm first to build a tree on an undirected graph from the MH, which corresponds to a downstream part of the entire multicast route, as shown by the blue dashed lines. Then, in Model 1, terminal paths are constructed to keep each of them as shortest as possible by adding unused links and reverse links to different terminal paths, as shown by the green dotted lines and the red lines, respectively. On the other hand, Model 2 tries to reduce the number of terminal paths by combining unused links and reverse links to increase the length of each terminal path. 
For example, at the Node 3 of Fig. \ref{p_route}, in Model 1, there are three different paths: two unused paths in the green dotted lines and one reverse path in the red line. Whereas, in Model 2, the reverse path from Node 3 to Node 2 is combined to the unused path from Node 5, resulting in two different paths at Node 3 in total. 
In general, with short terminal paths, both models operate robustly and accurately. However, Model 2 with less number of terminal paths has a better performance compared with Model 1. 

Both unicursal and our previously proposed route schemes have shortcomings because of their extremes. The unicursal route with maximum path length requires a smaller number of accesses to OFSs to locate all high-loss links but needs a larger number of probe packets to operate accurately. Our previous shortest path tree-based route scheme with shorter-length paths requires a smaller number of probe packets to keep an accuracy in loss conditions but needs a larger number of accesses to OFSs to locate all high-loss links. In both extremes, the number of and the lengths of terminal paths cannot be intentionally controlled. In the next subsection, we propose a new route scheme that can balance the number of terminal paths and their lengths, which is evaluated through simulation later in Section 6.

\subsection{The proposed backbone-and-branch tree route scheme (BBT)}
The newly proposed route scheme is called the backbone-and-branch tree route scheme (BBT) and its flowchart is illustrated in Fig. \ref{flowchart}. In BBT scheme, the Eulerian cycle algorithm is applied to the original undirected graph (network).
Since a Eulerian cycle only exists in the graph such that every vertex has an even degree, first we need to process all odd degree vertices (nodes) by temporarily omitting some links in general. Then we generate the backbone paths by using the Eulerian cycle algorithm to efficiently cover links as many as possible and also by considering to avoid too long terminal paths.
After generating the backbones, the reverse direction segments of route (i.e., toward the measurement node) on the backbone paths are added to the route tree at some branch nodes. 
Note that segment is a sequence of adjacent directed links to form a part of route.
Finally, we integrate paths of omitted links (called omitted paths) into the route tree. 

\begin{figure}[t]
	\centering{\includegraphics[width=0.65\linewidth]{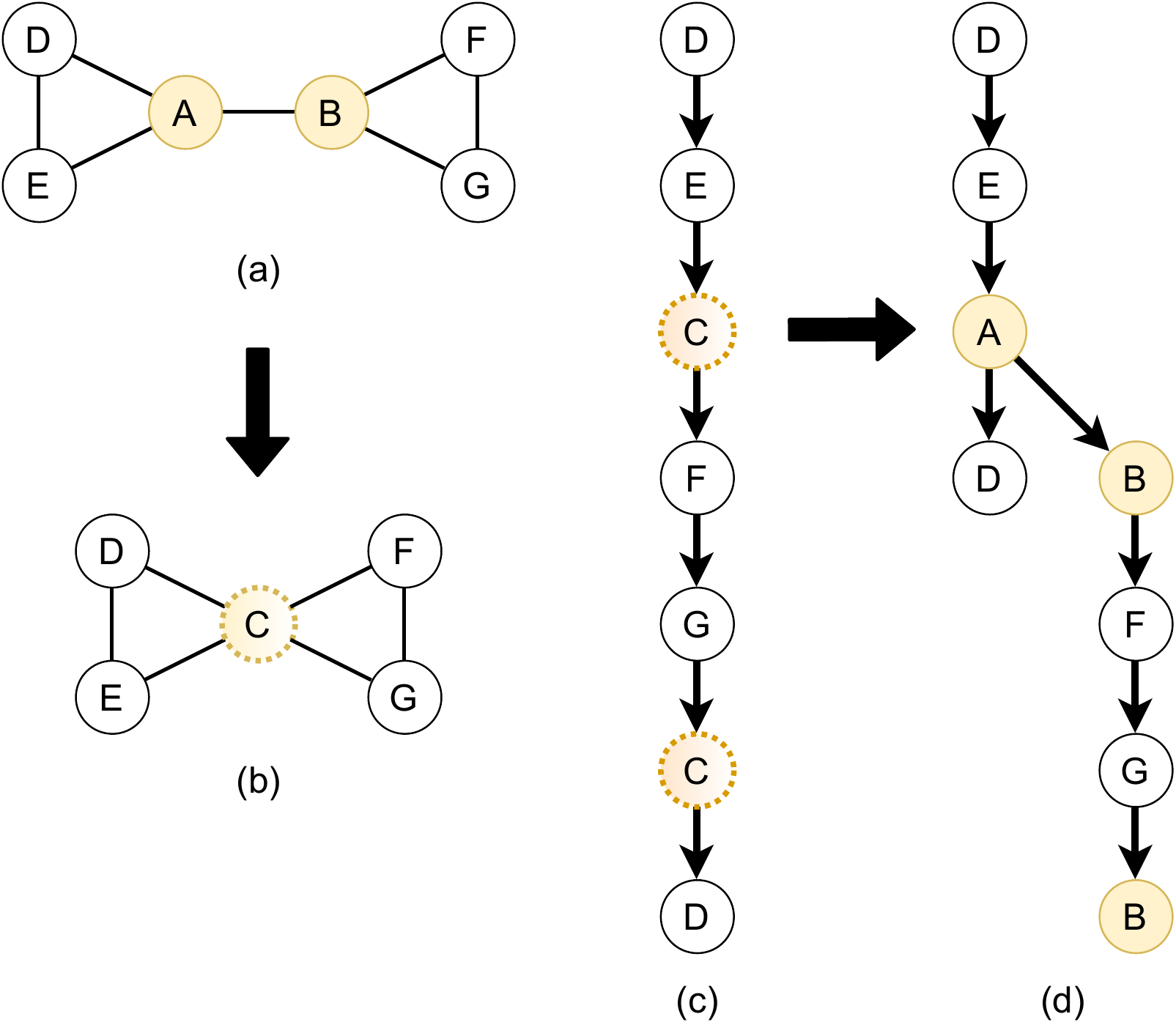}}
	\caption{Example of a cut path}
	\label{cutpath}
\end{figure}

\subsubsection{Omit links incident to odd degree nodes}
The general idea of processing odd degree nodes is omitting links between their couples. In this way, the degrees of these nodes become zero or even. Note that the number of odd degree nodes is always even.
Omitting links is based on the following criteria with the priority: (1) try to keep the network connected; and (2) minimize the number of omitted links.

If the network obtained by omitting links incident to nodes with odd degrees is still connected, a simple backbone path can be generated. Otherwise, if several omitted links become a cut set in the network (the network becomes disconnected if remove this path), the backbone path is fragmented and impact on the performance. In this case, the cut omitted path is replaced by a combined node in the network, see Fig. \ref{cutpath}\textbf{a}. The omitted path between Node A and Node B is combined into node C, the dotted node inFig. \ref{cutpath}b. From that, the network is still connected, and all nodes have even degree. Fig. \ref{cutpath}c is the backbone path from the combined network. However, the final backbone path in Fig. \ref{cutpath}d is fragmented. So, keeping the network connection is the highest priority when omitting odd degree nodes.

\begin{figure}[t]
	\centering{\includegraphics[width=0.6\linewidth]{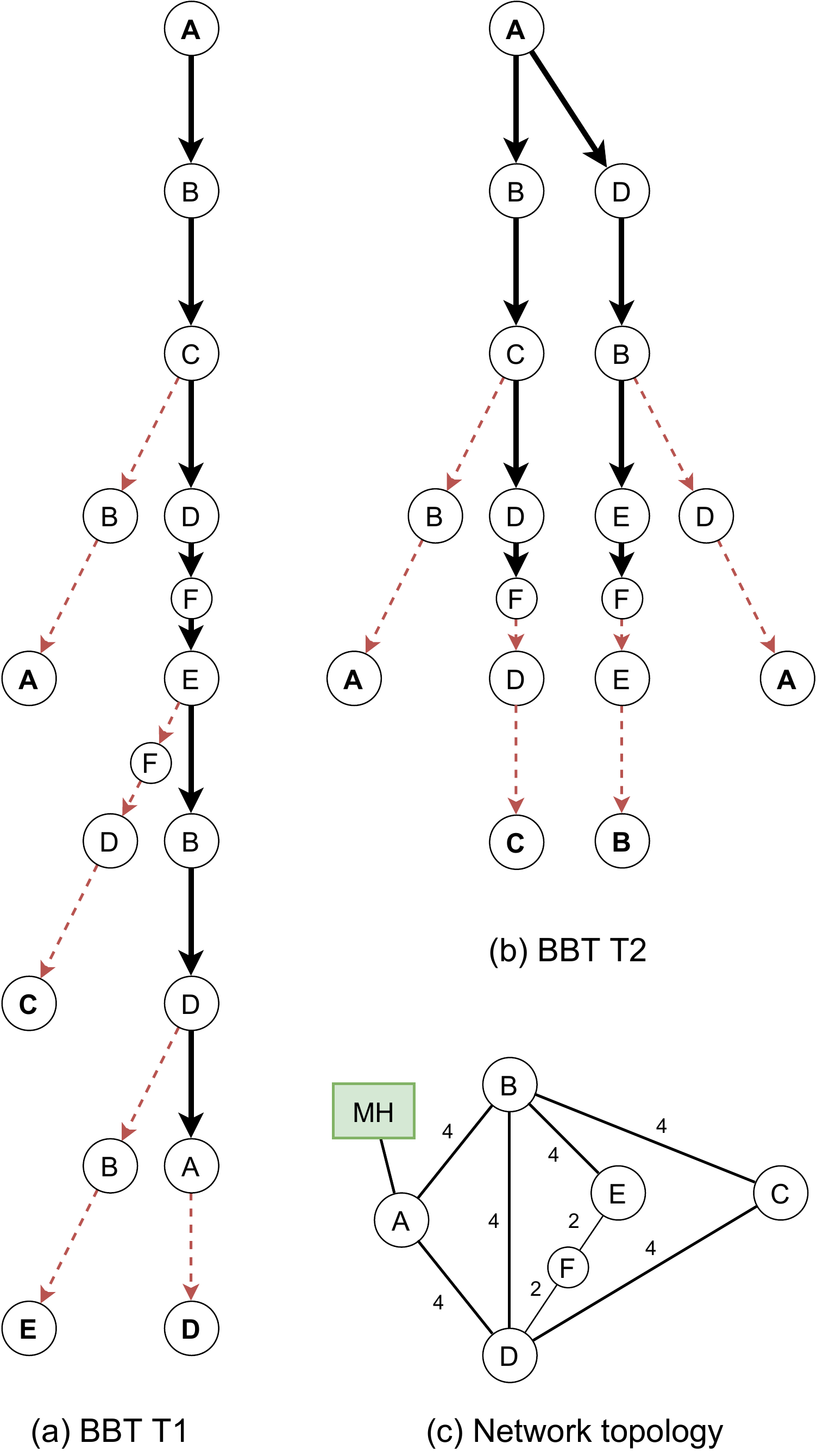}}
	\caption{Illustration of the backbone paths and the reverse branches}
	\label{euler}
\end{figure}

To minimize the omitted links, it takes many computations for searching all combinations of each odd degree couple. Therefore, we propose an approach to reduce the computation and the time by omitting in the following sequence. Firstly, we omit links of the unique path between a 1-degree node and another odd degree node. Secondly, the link between two neighbor odd degree nodes is omitted.
Finally, from the remaining odd degree nodes, we calculate the shortest paths from each node to the others then select the optimal combination. 
The selected combination is the case that the number of  the omitted links is minimized and the network is still connected.
Each series of omitted links is called an omitted group.
Note that if two omitted groups have a common node, they are combined into one omitted group.

\subsubsection{Generate the route tree}

After the omitting step, each node in the network has an even degree. We can generate the backbone paths of probe packets. Then, for each segment of the backbone path, its reverse path is added as the branches of the route tree. 
A reverse path or reverse segment is the reverse direction segment of route toward the measurement node on the backbone path. 
This process is illustrated in Fig. \ref{euler}. Fig. \ref{euler}c is an artificial Eulerian cycle-based network topology without odd degree nodes. The numbers on links in the figure indicate the link distance between two nodes.
In other words, there are other unnamed nodes/links between those two nodes that are counted in the length of paths and segments, while they are not explicitly illustrated in the figure.
From node A as the measurement node, the node that connects to the measurement host, the backbone path of the route tree is generated based on the Eulerian cycle algorithm. Here, there are two options: 1 backbone path case (BBT T1) or 2 backbone paths case (BBT T2). BBT T1 has a full Eulerian cycle as the backbone path, the bold line as in Fig. \ref{euler}a. The path starts from node A, travels through nodes B, C, D, F, E, B, D, and ends at node A (also the start node). BBT T2 has two backbone paths that are two reversed halves of the Eulerian cycle, see Fig. \ref{euler}b. They are the A-B-C-D-F path and the A-D-B-E-F path, which meet at node F in the middle of the Eulerian cycle. This option can halve the length of the backbone path.

\begin{figure}[t]
	\centering{\includegraphics[width=0.7\linewidth]{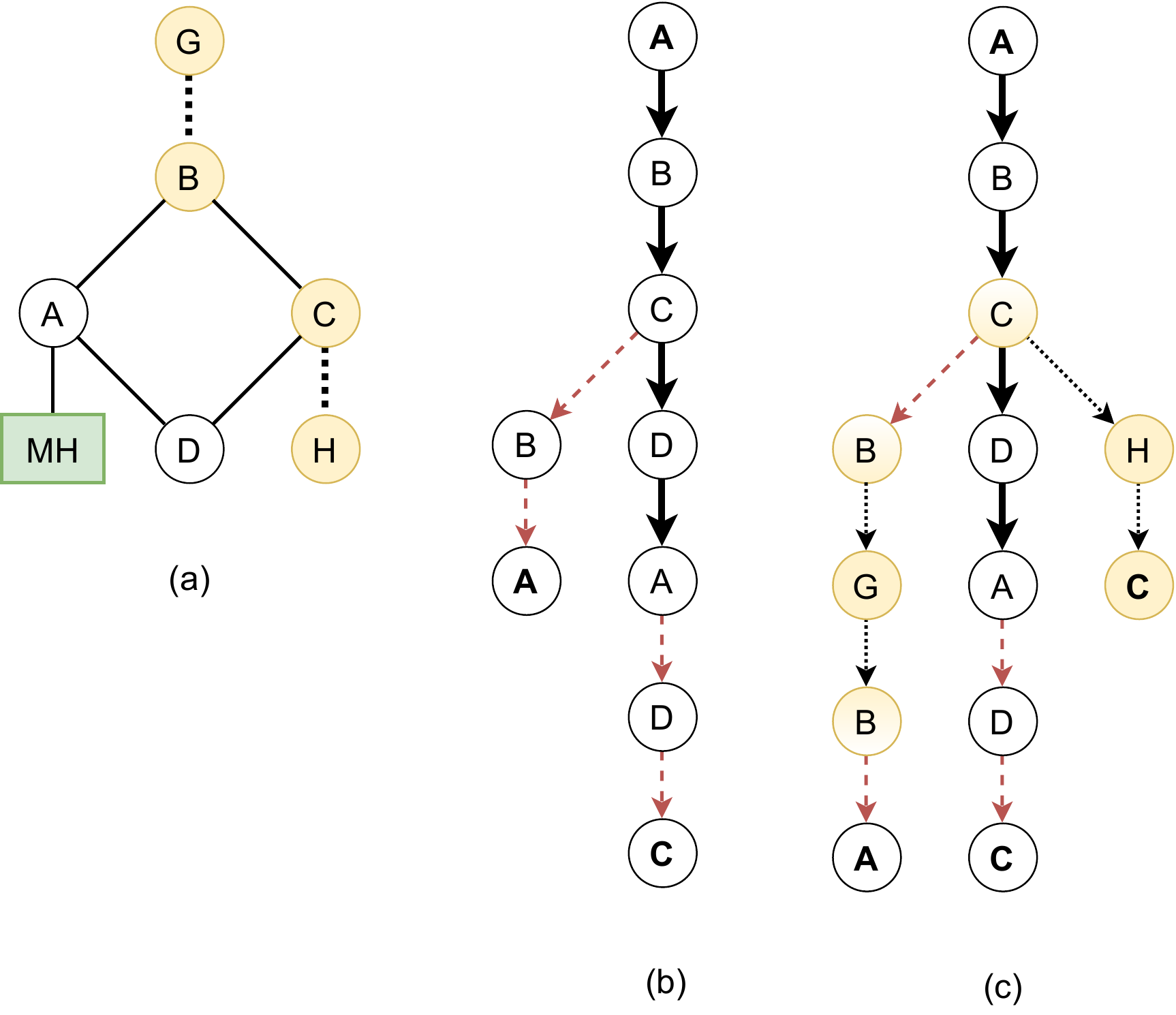}}
	\caption{Example of integrating omitted paths}
	\label{addpath}
\end{figure}

Note that, the number of backbone paths directly affects the number of and the lengths of terminal paths of measurement route, essentially related with the trade-off between the tolerance to heavy loss of measurement packets and the efficiency in lossy link detection by measurement packets.
We should properly choose the number of backbone paths depending on the expected loss rates on links, that is the degree of lossiness of links.
In the proposed scheme, the number of backbone paths is limited by the degree of measurement node.
If the degree of measurement node is two, we can construct two backbone paths at maximum. This is the example situation used in this paper.
If the degree of measurement node is more than two, although it was not quantitatively evaluated, we can construct more than two backbone paths. For example, in Fig. \ref{euler}c, if the measurement node is Node B, three backbone paths can be constructed. They are the B-A-D, B-C-D, and B-D-F-E paths.
However, it is desirable to more flexibly decide the number of backbone paths in order to control the above trade-off in response to the network topology and link conditions. 
A possible option to increase such flexibility is to consider multiple MHs in either physical or virtual manner.
For example, an appropriate set of nodes can be selected as virtual measurement nodes from which a backbone path will start. In this setting, the probe packets should be forwarded from a real MH to each of virtual measurement nodes along some nodes/links.
However, it remains as future work.

To build reverse paths, first, we divide a backbone path into multiple backbone segments. The length of each backbone segment decides the number of backbone segments (it is also the number of terminal paths). As we will discuss later in Section 5, it is desirable that the length of segment is similar across all segments.
The segmentation should take this point into consideration.
Then, at the end node of each segment, called branch node, the reverse path is added, like dashed lines in Fig. \ref{euler}.
Each reverse path has the same length but opposite direction with its backbone segment. For example, in Fig. \ref{euler}a, there are four segments: A-B-C, C-D-F-E, E-B-D, and D-A on the backbone. So, 4 reverse paths (segments: C-B-A, E-F-D-C, D-B-E, and A-D) are added to the route tree and generate 4 terminal paths.

After adding reverse paths, the omitted paths are integrated into the route tree. From the common node of the generated route tree and an omitted group, a unicursal path (a single route visits all links of the omitted group in both directions) is generated. This path is integrated into the reverse path or added to the backbone path as an individual branch path.
For an example in Fig. \ref{addpath}, the route tree before integrating omitted paths is in Fig. \ref{addpath}b. With the omitted group B-G, the common node is Node B, in the middle of the reverse path. So, the omitted path B-G-B is integrated into the reverse path C-B-A and this reverse path becomes the integrated reverse path C-B-G-B-A, see the left branch in Fig. \ref{addpath}c. 
To prevent a path from becoming too long, the omitted path can be added to the backbone path as an individual branch path if the common node is the branch node, see the right branch path C-H-C in Fig. \ref{addpath}c.
In the complete route scheme, all paths from branch nodes (including reverse paths, integrated reverse paths, and individual branch paths) are called branch paths or branch segments. If there is only one branch path from the final branch node (the end node of the final backbone segment), this branch path is included the final backbone segment.

\begin{figure}[t]
	\centering{\includegraphics[width=0.65\linewidth]{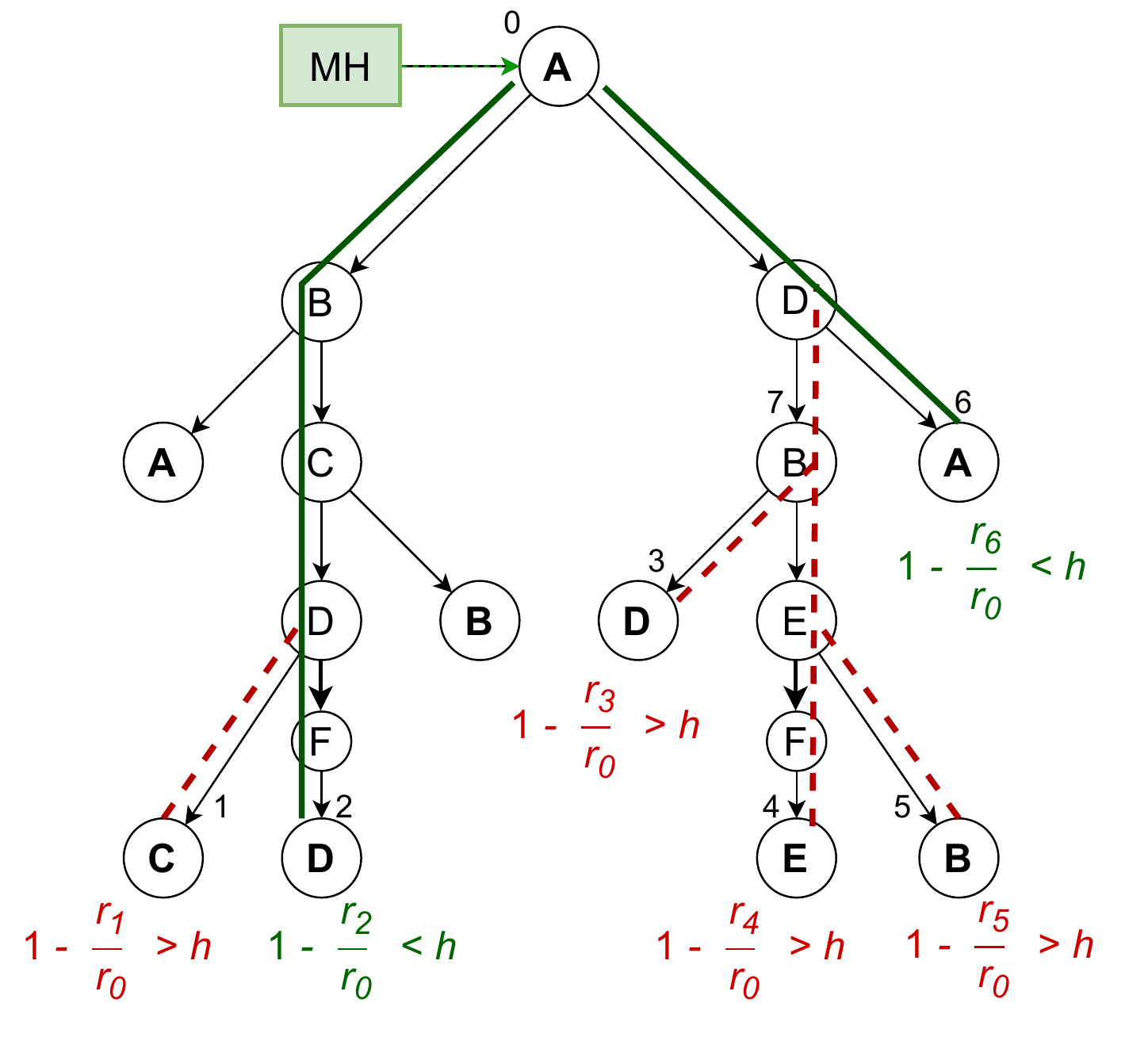}}
	\caption{Example of the access order}
	\label{order}
\end{figure}

Note that the Eulerian cycle can start from any node. If a given measurement node is omitted in the ``omit links incident to odd degree nodes" step, the nearest remaining node is considered as the new measurement node. A segment of this omitted path from the original MH to the new MH is integrated into the backbone path, and the rest is the branch path.

\section{Sequence Access Order}

After receiving the probing complete notification from the MH, the OFC begins the next
process to locate high-loss links by querying and collecting flow statistics information from selected OFSs in an appropriate access order. Note that link is considered as high-loss if and only if its loss rate exceeds threshold value $h$. Here, $h$ is a design parameter that represents the target link quality to be maintained, which depends on the target applications. The packet loss rate (PLR) of a range (a sequence of links) from ports $i$ to $j$, $PLR = 1 - \frac{r_j}{r_i}$, where $r_i$ and $r_j$ are the numbers of probe packets arriving at switch ports $i$ and $j$ , respectively.

First, the OFC calculates the PLR of each terminal path with the information at the root port and leaf ports. If the PLR of a terminal path is less than $h$, this terminal path does not include a high-loss link. Otherwise, this terminal path is likely to include one or more high-loss link. Then, by considering the correlation among terminal paths in terms of the degree of packet loss, we can narrow the search scope, i.e., the expected locations of high-loss links. 
If a terminal path is high-loss and there are no other high-loss terminal paths, the high-loss links are located within a range between the leaf port and the nearest branch port on the considered high-loss terminal path. The dashed line on the left part in Fig. \ref{order} shows an example of this case. The binary search algorithm is used to locate all loss links in a range.

If there are multiple terminal paths whose PLR values exceed threshold $h$, the port most commonly shared by those paths is queried first to collect the number of probe packets that have arrived, which produces separated subtrees, and the same procedure is performed recursively. An example of this case is illustrated in the right part of Fig. \ref{order}, the next requested port is the port $7$ of node B. The actual packet loss rate of each high-loss link is measured exactly based on the difference between the numbers of arriving probe packets at the link’s upper and lower ports.

\section{Discussion}

\begin{figure}[t]
	\centering{\includegraphics[width=1\linewidth]{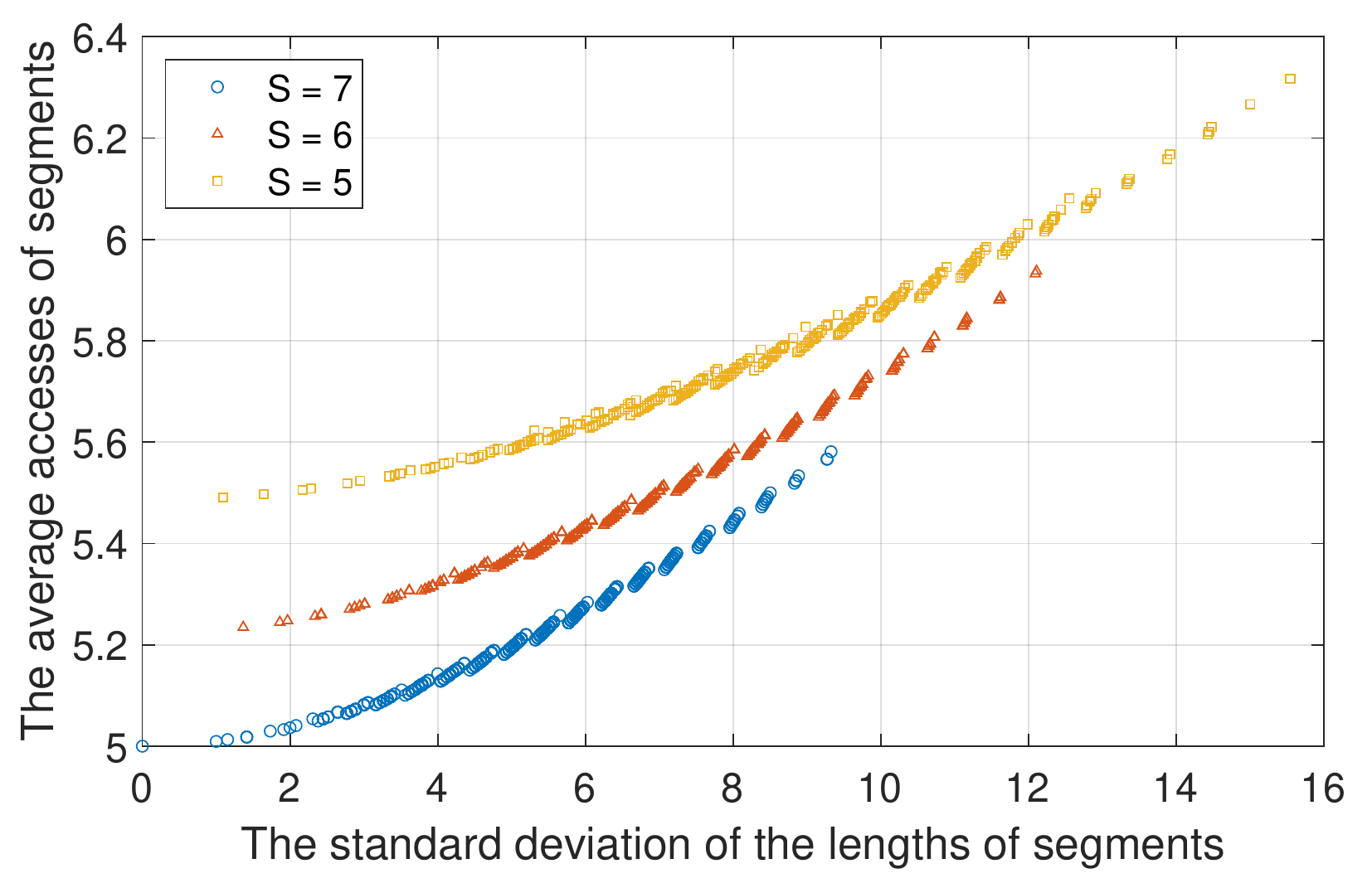}}
	\caption{The average accesses of segments}
	\label{varlen}
\end{figure}

In this section, we will discuss issues when designing the route scheme that affect the performance of the measurement. They are the length of segments and related problems such as the difference in length of segments, impacts of omitted paths and cut paths on the route scheme.

Assume that there is only a high-loss link in a given segment, the average number of required accesses of a segment is as follows. If the length $s$ of the segment is $2^k$, $k = 1, 2, 3,...$

\begin{equation}
	F_{seg} = 2 + k.
\end{equation}
Otherwise,

\begin{equation}
	F_{seg} \approx 2 + \log_2 s.
\end{equation}
The OFC requests the first and the last ports of this segment. Then, the binary search algorithm is used to locate the high-loss link. 

Let $S$ be the number of segments.
We number (as index) all the segment from 1 to $S$ so that 
the index of the first common segment is 1 and
the indexes of the branch segments are $S-B+1, \ldots, S-1, S$,
respectively. $B$ is also the number of terminal paths.
Let $s_i$ be the length of the $i^{th}$ segment, and $n$ be the number of links.
If the single high-loss link is randomly located
at any link with the same probability $1/n$,
the average number of accesses to locate the high-loss link is

\begin{eqnarray}  \label{ft1}
	F_{T1} \approx 1 + B + \frac{s_1}{n}(1 + \log_2 s_1)
	+ \sum_{i=2}^{S-B} \frac{s_i}{n}(2 + \log_2 s_i)  \nonumber \\ 
	+ \sum_{i=S-B+1}^{S} \frac{s_i}{n}(1 + \log_2 s_i). 
\end{eqnarray}

The number of accesses is the sum of accesses of one root port, $B$ leaf ports, the average accesses of the segment which has the high-loss link. Since the OFC requested the root port and all leaf ports, when searching on the first common segment and branch segments, only one port is needed to request.

\begin{table}[tb]
	\caption{The number of terminal paths and the path length of route schemes on the ideal topology}
	\label{len:ideal}
	\hbox to\hsize{\hfil
		\begin{tabular}{l|llll}\hline
			&Paths &Average &Min &Max\\\hline
			Unicursal& 1 & 56 & 56 & 56 \\
			T1\_seg8&	4 & 26 & 16 & 32\\
			T2\_seg8&	4& 18 & 16 & 20 \\
			T1\_seg4&	7 & 20 & 8 & 32\\
			T2\_seg4&	8 & 13 & 8 & 16 \\\hline
			\multicolumn{5}{l} {Paths: The number of terminal paths\quad}\\
			\multicolumn{5}{l} {Average: The average length of terminal paths.}\\
			\multicolumn{5}{l} {Min: the minimum length. \quad
				Max: the maximum length.}\\
		\end{tabular}\hfil}
\end{table}

Since each probe packet traverses each link only once, the total length of segments is the number of links. The length of each segment impacts on the number of accesses. Fig. \ref{varlen} shows the relationship between the total of the average accesses of segments and the standard deviation of the lengths of segments. The number $n$ of links is $56$, and the numbers of segments are 5, 6, and 7. The optimal values happen at small the standard deviations. That means the length of segment is better to be similar across all segments. Note that omitted paths are added to branch segments, so minimizing omitted paths length keeps the difference between segments small. Additionally, because of using binary search to locate high-loss links, the segment length with the $2^k$ form is a better choice.

If the omitted path is a cut path, besides generating new branch path (causing more accesses), the difference in length among segments also becomes large; and thus the number of accesses becomes large. Thus, keeping the network connected in the backbone is important to give a good performance.

On the other hand, the length of the segment also decides the number of terminal paths.
The shorter segment length is, the more terminal paths are generated, see Table \ref{len:ideal}. In the table, T1\_seg$x$ is the BBT T1 route scheme with the segment length is $x$, and T2\_seg$x$ is similar to T1\_seg$x$.
An acceptable (maximum) length of terminal paths is determined by the degree of lossiness of links on the network.
In addition, if the length of a segment is long and/or the loss rates of lightly lossy links on the segment is not very low, the measured loss rate of the segment may exceed a threshold of high-loss link, which results in more number of accesses due to a miss decision in narrowing high-loss ranges and finally locating high-loss links.
Since the length of segment is better to be similar across all segments, the length of a segment on the backbone path is a parameter of our route design. 
However, an appropriate length of each segment depends on the lossiness of the links, which may be predicted (estimated) based on the past information by constantly monitoring the network. However, this issue remains as our future work.
In general, for massive loss networks, the length of each segment should be short and vice versa. For the evaluation in Section 6, we examine route schemes in different loss environments.

\section{Evaluation}

\begin{figure}[t]
	\centering{\includegraphics[width=0.9\linewidth]{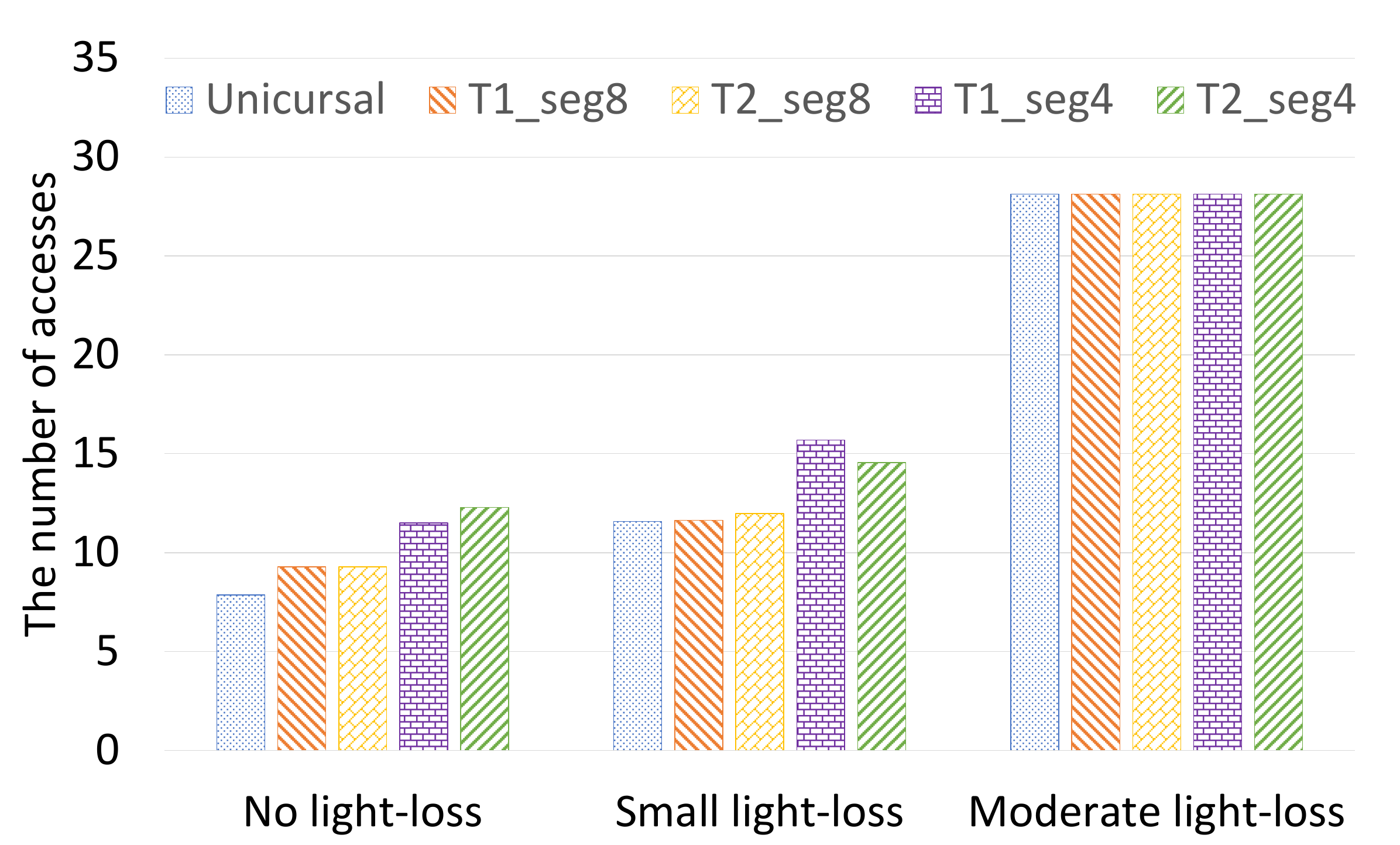}}
	\caption{The number of accesses to locate 1 high-loss link on the ideal topology}
	\label{perftree1}
\end{figure}

\begin{figure}[t]
	\centering{\includegraphics[width=1\linewidth]{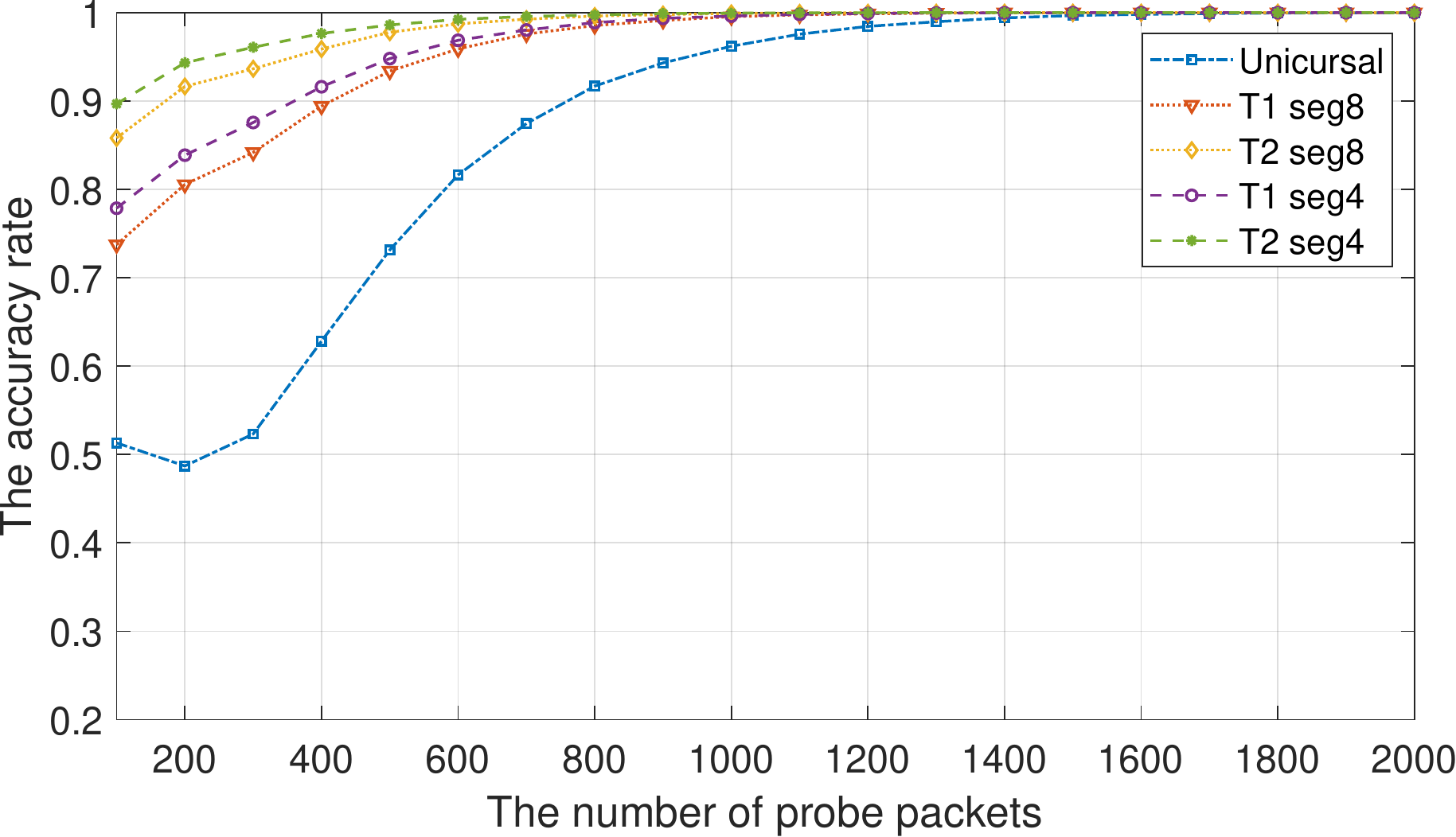}}
	\caption{The accuracy rate of the measurement in the moderate light-loss environment}
	\label{acc}
\end{figure}

To evaluate the performance of route schemes, we consider the number of required accesses of the OFC to OFSs. First, we use an artificial ideal topology, Fig. \ref{euler}c, to compare the performance of different route schemes and their segment length. In this topology, the link distance between two nodes is 4; this
means there are 4 links and 3 hidden nodes between them, excluding
the D-F link, and the E-F link is 2. We assume high-loss links have loss rates from 0.15 to 0.2, the threshold is 0.1. Other links are considered in three cases: (1) No light-loss, (2) have small light-loss rates with the value from a range of [0, 0.02], and (3) have moderate light-loss rate with the value from a range of [0, 0.04]. The number of probe packets is 100,000. The samples are 100,000. The information of terminal paths of route schemes is in Table \ref{len:ideal}.

Fig. \ref{perftree1} shows that in small loss or no loss environments, the route scheme with less number of terminal paths has better performance. 
In the massive loss environment, the performance of route schemes is similar, see the third column group in Fig. \ref{perftree1}. In this case, although there is no high-loss link, the accumulated loss rate of links in terminal paths can exceed the threshold, resulting in more accesses to locate actual high-loss links. 

The results in Fig. \ref{acc} show the relationship between the accuracy and the number of probe packets in the mass loss network with 2 high-loss links with the loss rate that is randomly selected from a range of [0.5, 0.7], other links have moderate light-loss rates. We see that the route scheme with longer path length needs more probe packets to operate accurately. In comparison with the unicursal route length, the longest path of our proposal is about $50\%$ in the BBT T1 and $25\%$ in the BBT T2. Besides, by reducing the segment length, we can generate more terminal paths with shorter lengths and increase the accuracy. 

\begin{figure}[t]
	\centering{\includegraphics[width=0.85\linewidth]{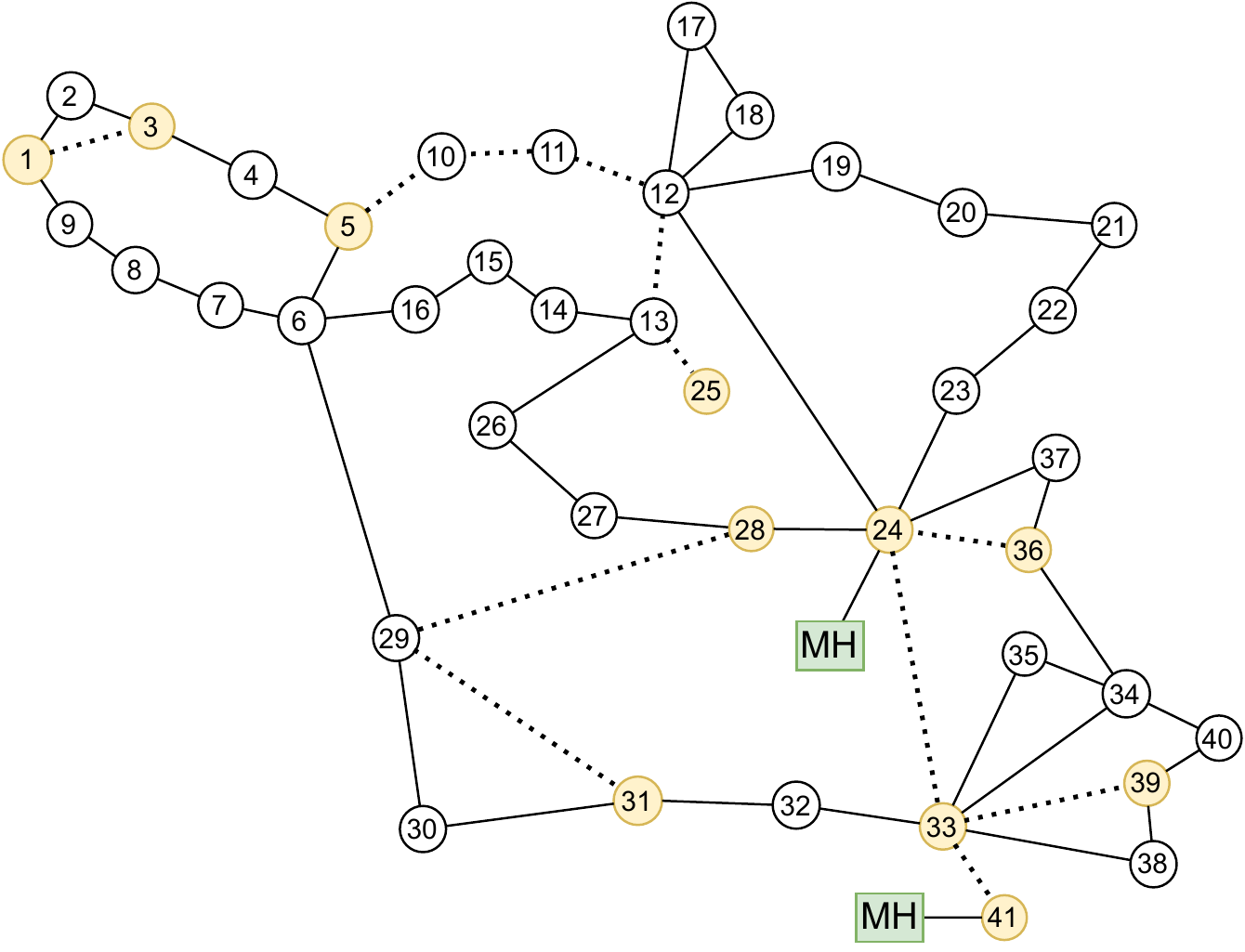}}
	\caption{The Renater network topology \cite{topozoo}}
	\label{renater}
\end{figure}

\begin{table}[tb]
	\caption{The information of terminal paths and segments of route schemes on the Renater topology}
	\label{len:renater}
	\hbox to\hsize{\hfil
		\begin{tabular}{l|llllll}\hline
			&Paths &Average &Min &Max &S &stdev\\\hline
			Unicursal& 1 & 108 & 108 & 108&1&0 \\
			T2\_seg8\_MH@24&	8 & 19.625 & 8 & 28&13&2.51\\
			Model 2\_MH@24&	26& 5.5 & 3 & 12 &--&--\\
			T2\_seg8\_MH@41&	8 & 21.125 & 8 & 29 &14&3.14\\
			Model 2\_MH@41&	25 & 7 & 5 & 14&-- &--\\\hline
			\multicolumn{7}{l} {S: the number of segments}\\
			\multicolumn{7}{l} {stdev: the standard deviation of the lengths of segments}\\
		\end{tabular}\hfil}
\end{table}

To evaluate the performance in the large-scale network, we consider a network topology based on the real network, illustrated in Fig. \ref{renater}. In this simulation, the number of high losses is changed from 1 to 4, with the loss rate is from 0.15 to 0.2. Other links have a random light loss value from a range of [0, 0.02]. The number of probe packets is 100,000. The threshold is 0.1. The BBT T2 is used with 2 backbone paths, and the segment length is 8. Dotted lines show omitted paths. Two situations of the MH location, at Node 24 and Node 41, are considered.

\begin{figure}[t]
	\centering{\includegraphics[width=1\linewidth]{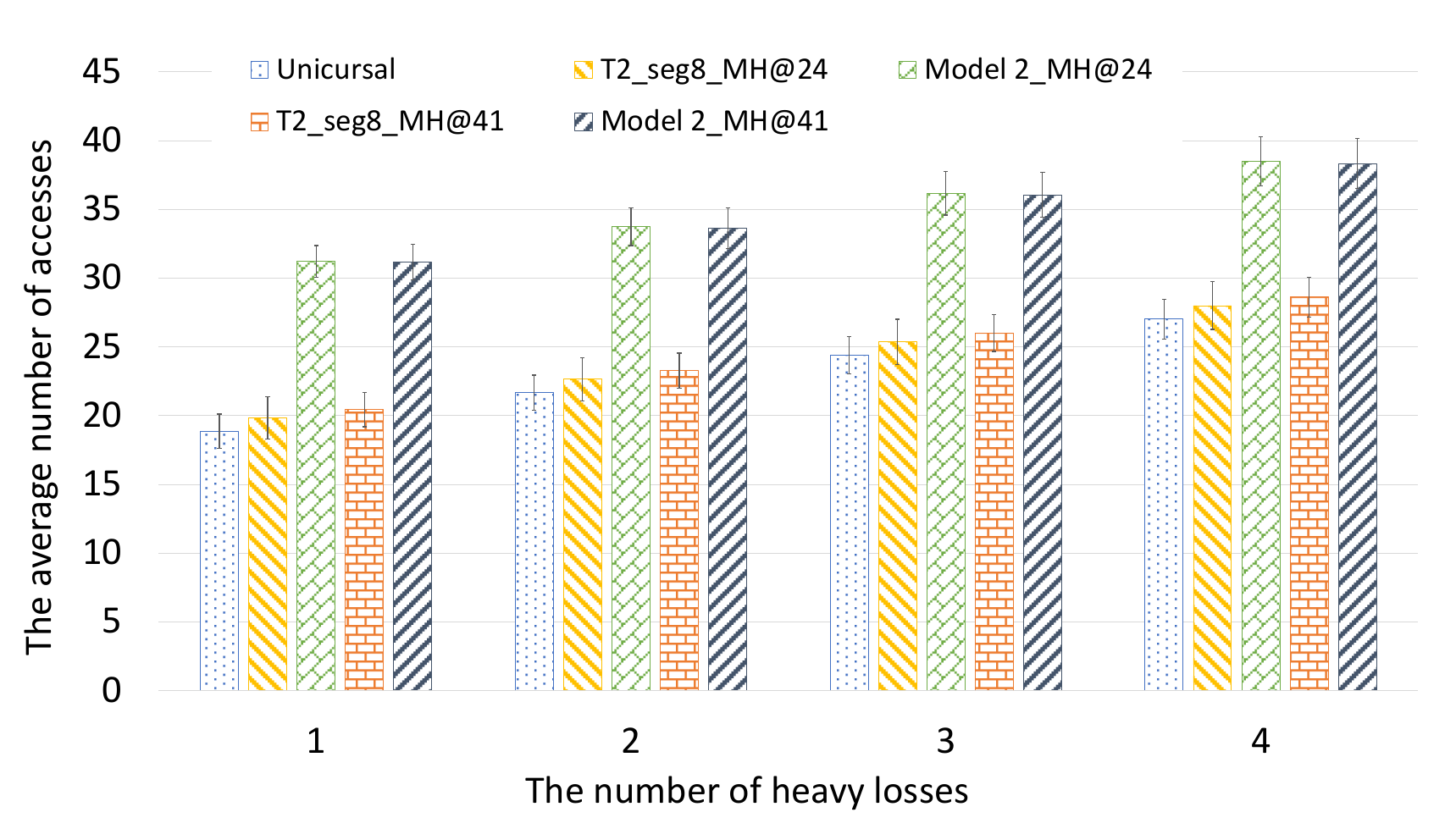}}
	\caption{The number of required accesses to locate high-loss links on the Renater topology in the small light-loss environment}
	\label{perf}
\end{figure}

\begin{figure}[t]
	\centering{\includegraphics[width=1\linewidth]{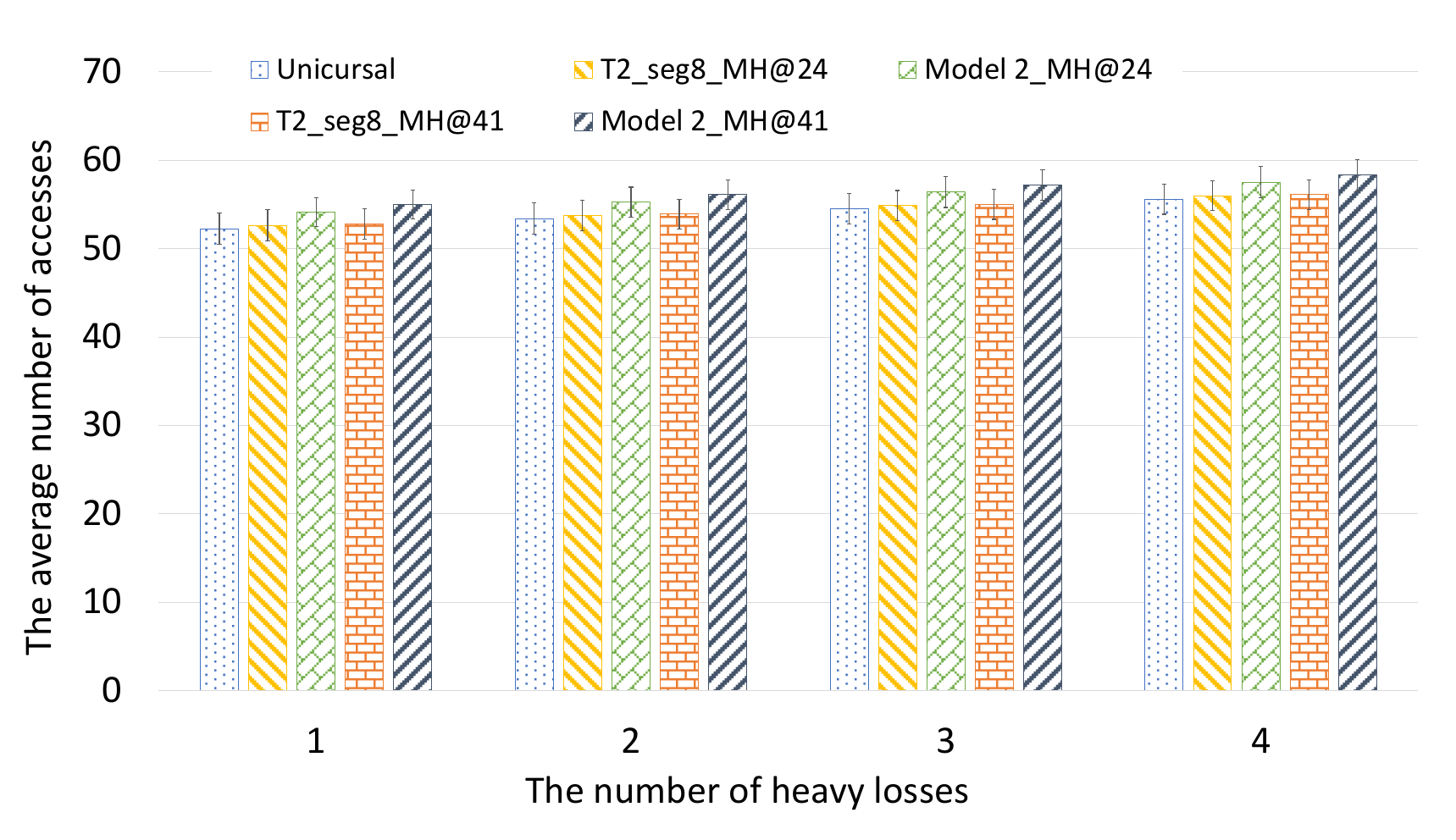}}
	\caption{The number of required accesses to locate high-loss links on the Renater topology in the moderate light-loss environment}
	\label{perf2}
\end{figure}

\begin{figure}[t]
	\centering{\includegraphics[width=1\linewidth]{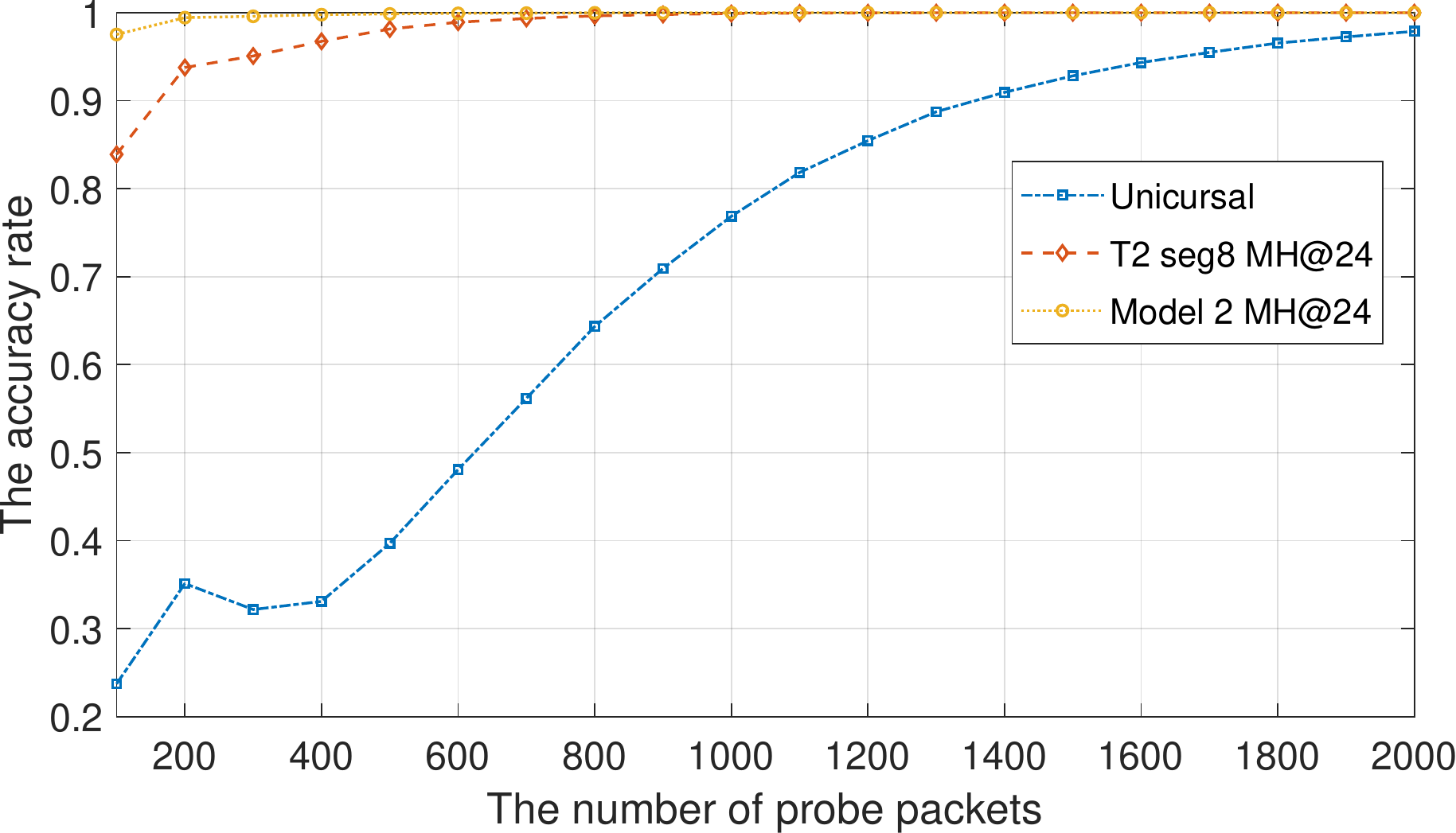}}
	\caption{The accuracy rate on the Renater topology in the moderate light-loss environment}
	\label{re_acc}
\end{figure}

Fig. \ref{perf} shows the number of required accesses on the Renater topology with different methods and MH locations. Our proposal has the performance close to the unicursal route and improves significantly compared to the heuristic variant model (Model 2 in \cite{TriISCC19}) in the small light-loss environment. In the moderate light-loss environment, the number of accesses of our proposal is also smaller, see Fig. \ref{perf2}. In this environment, the number of probe packets affects the accuracy remarkably. Fig. \ref{re_acc} shows the relationship between the number of probe packets and the accuracy of the measurement. Our proposal also has the accuracy rate close to the previous method. 

Note that, in an ideal topology without odd degree node, our proposed route scheme is independent of the location of the MH. However, in a general network, the location of the MH has a certain influence, especially at the omitted nodes. This shows in the performance of two cases with the MH connects to the omitted node, Node 41, and the remaining node, Node 24. The case with the MH at the omitted node is worse. This is because the omitted path included the MH will be fragment into 2 parts. One part is integrated into the backbone path, the other is the branch path. So the route scheme, in this case, have more segments, and the standard deviation of the lengths of segments is also lager than the case with the MH at remaining Node 24, see Table \ref{len:renater}.

In conclusion, the simulation results show the effectiveness of the proposed BBT route scheme in our framework for an ideal network as well as a real network. 
It can keep the terminal paths short enough to avoid errors in loss rate estimation compared to the unicursal route, while keeping the number of terminal paths small enough to reduce the number of accesses to switches significantly compared to the previously proposed shortest-path tree based route.

\section{Concluding Remarks}
We have proposed a practical framework of monitoring both directions of all full-duplex links of an entire OpenFlow-based network to promptly locate high-loss links with a minimized load on both data-plane and control-plane incurred by the measurement.
The framework introduces a combination of an active measurement by probing multicast packets along a designed route and a passive measurement by collecting flow-stats of the probing flow at selected switch ports in an appropriate sequential order to access switches.

In particular, we have proposed the BBT route scheme. Through simulation-based evaluations on a real-world large network topology, it was shown to balance between the measurement accuracy and the measurement overhead (i.e., the number of accesses to switches until locating all high-loss links).
Note that our framework was implemented in the Ryu OpenFlow framework and tested on a Mininet environment; and is planned to test on a nation-wide OpenFlow testbed.

As future work, we should challenge to adaptively optimize the schemes of the multicast probe packet route for flowing on every link and the access order to switches for collecting flow-stats by reflecting the past measurement results in continuous monitoring scenarios, e.g., by leveraging machine learning techniques.

\end{document}